\documentclass{article}

\PassOptionsToPackage{numbers, compress}{natbib}
\usepackage[preprint]{neurips_2023}

\usepackage[utf8]{inputenc} 
\usepackage[T1]{fontenc}    
\usepackage{hyperref}       
\usepackage{url}            
\usepackage{booktabs}       
\usepackage{amsfonts}       
\usepackage{nicefrac}       
\usepackage{microtype}      
\usepackage{xcolor}         
\usepackage{natbib}
\usepackage{multirow}
\usepackage[normalem]{ulem}
\useunder{\uline}{\ul}{}
\usepackage{colortbl}
\usepackage{amsmath}
\usepackage{grffile}
\usepackage{graphicx}

\definecolor{mygreen}{rgb}{0.0, 0.5, 0.0}

\newcommand{\revise}[1]{{\color{black}{#1}}}

\newcommand{\ea}{\textit{et al.}}
\newcommand{\ourapp}{\textsc{OptiMatch}}

\title{Learning to Quantize Vulnerability Patterns and Match to Locate Statement-Level Vulnerabilities}

\author{
Michael Fu$^1$~~~
Trung Le$^1$~~~
Van Nguyen$^{1,2}$~~~ \\
\textbf{
Chakkrit Tantithamthavorn$^1$~~~
Dinh Phung$^{1,3}$~~~
}
\smallskip
\\
$^1$Monash University, Australia
\\
$^2$CSIRO’s Data61, Australia
\\
$^3$VinAI Research, Vietnam
\\
}

\begin{document}

\maketitle

\begin{abstract}
Deep learning (DL) models have become increasingly popular in identifying software vulnerabilities. \revise{Prior studies found that vulnerabilities across different vulnerable programs may exhibit similar vulnerable scopes, implicitly forming discernible vulnerability patterns that can be learned by DL models through supervised training.} However, vulnerable scopes still manifest in various spatial locations and formats within a program, posing challenges for models to accurately identify vulnerable statements. Despite this challenge, state-of-the-art vulnerability detection approaches fail to exploit the vulnerability patterns that arise in vulnerable programs. To take full advantage of vulnerability patterns and unleash the ability of DL models, we propose a novel vulnerability-matching approach in this paper, drawing inspiration from program analysis tools that locate vulnerabilities based on pre-defined patterns. Specifically, a vulnerability codebook is learned, which consists of quantized vectors representing various vulnerability patterns. During inference, the codebook is iterated to match all learned patterns and predict the presence of potential vulnerabilities within a given program. Our approach was extensively evaluated on a real-world dataset comprising more than 188,000 C/C++ functions. The evaluation results show that our approach achieves an F1-score of 94\% (6\% higher than the previous best) and 82\% (19\% higher than the previous best) for function and statement-level vulnerability identification, respectively. These substantial enhancements highlight the effectiveness of our approach to identifying vulnerabilities. The training code and pre-trained models are available at \url{https://github.com/optimatch/optimatch}.
\end{abstract}

\vspace{-3mm}
\section{Introduction}
\vspace{-2mm}
The number of software vulnerabilities has been escalating rapidly in recent years.
In particular, National Vulnerability Database (NVD)~\cite{nvd} reported 26,448 software vulnerabilities in 2022, soaring 40\% from 18,938 in 2019.
The extensive use of open-source libraries, in particular, may contribute to this rise in vulnerabilities.
For instance, the Apache Struts vulnerabilities~\cite{apache} indicate that this poses a tangible threat to organizations.
The root cause of these vulnerabilities is often insecure coding practices, making the source code exploitable by attackers who can use them to infiltrate software systems and cause considerable financial and social harm.

To mitigate security threats, security experts leverage static analysis tools that check the code against a set of known patterns of insecure or vulnerable code, such as buffer overflow vulnerabilities and other common security flaws.
In contrast, deep learning-based vulnerability detection (VD) identifies vulnerabilities at the file or function levels by implicitly learning vulnerability patterns during training~\cite{russell2018automated,zhou2019devign,nguyen2021information,nguyen2022regvd}.
DL-based vulnerability detection (VD) methods have demonstrated higher accuracy compared to static analysis tools that only target specific vulnerability types~\cite{li2021sysevr,fu2022linevul,croft2021empirical}. Additionally, recent advancements have introduced fine-grained VDs that offer statement-level vulnerability predictions, aiming to minimize the manual analysis burden on security analysts.
Previous studies have employed graph structure of source code like the code property graph~\cite{cpg}, along with graph neural networks to detect vulnerabilities at the statement level~\cite{li2021vulnerability,hin2022linevd}. Additionally, transformers have demonstrated their capability to learn semantic features of code using self-attention, which is particularly beneficial for handling long sequences compared to RNN models~\cite{fu2022linevul,ding2022velvet}.

\begin{figure*}[h!]
\centering
\includegraphics[width=0.98\textwidth]{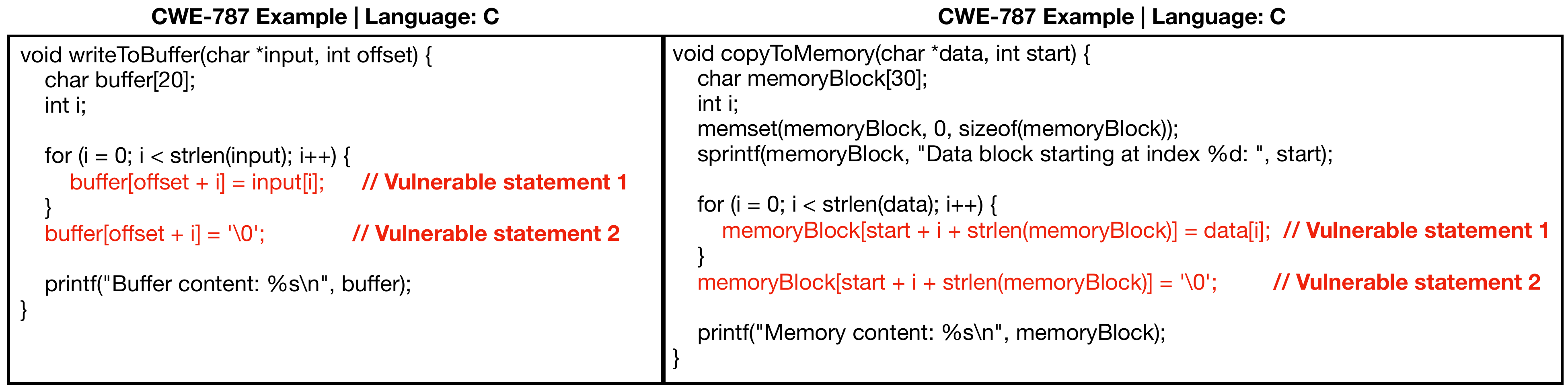}
\caption[Caption]{In the left function, \textit{writeToBuffer}, if the sum of \textit{offset} and \textit{i} exceeds or equals 20, it results in writing data beyond the buffer array's end. This overwrites memory beyond the array, posing a potential program crash. Similarly, the \textit{copyToMemory} function on the right uses the \textit{start} index to determine the starting point for copying data in \textit{memoryBlock}. However, if the sum of \textit{start} and \textit{i} surpasses or equals the size of \textit{memoryBlock}, it leads to overwriting memory beyond the array, causing an out-of-bounds write vulnerability. Despite sharing the same vulnerability type and similar vulnerable scopes, the vulnerable statements in each function are different in their written form, variable names, and positions.}
\vspace{-6mm}
\label{fig:example_code}
\end{figure*}

\revise{\textit{In this paper, we consider a vulnerable scope of a function as the collection of all vulnerable statements in that function.}
As illustrated in Figure~\ref{fig:example_code}, each function consists of two vulnerable statements that form similar vulnerable scopes. 
This suggests that even if two functions contain the same CWE-787 out-of-bound write vulnerability (the top-1 dangerous CWE-ID in 2022~\cite{top_25}), the specific vulnerable statements can be written in different ways and located in different parts of the code. Therefore, identifying vulnerabilities at the statement level is challenging for both machine learning and deep learning models. Despite this difficulty, our analysis reveals that state-of-the-art VD approaches have not successfully leveraged the information contained in vulnerable statements (that could be grouped to form vulnerable scopes) to further improve the capability of machine learning and deep learning vulnerability detection approaches at both the function and statement levels.

To address this issue, we propose a novel DL-based framework that can effectively utilize the information presented in vulnerable scopes. To achieve this, we develop a method for quantizing similar vulnerable scopes that share the same pattern into a vulnerability codebook consisting of common codewords which represent common patterns. This codebook captures a diverse range of vulnerabilities from the training data and facilitates the process of vulnerability matching inspired by the pattern-matching concept utilized in program analysis tools~\cite{checkmarx,cppcheck,flawfinder,rats}. Our approach is \textit{the first to successfully exploit the benefits of vulnerability matching and codebook-based quantization to enhance DL-based VD}. This allows us to effectively identify vulnerabilities in source code data, ultimately improving the overall capability of DL-based VD.

Our approach involves collecting and quantizing a set of vulnerable scopes from the training set before using the optimal transport (OT)~\cite{feydy2019interpolating} to cluster this set into a vulnerability codebook consisting of a set of vulnerability centroids (i.e., codewords).
The vulnerable scopes (collected from the training set) that share a similar pattern would stay closely in representations, hence we cluster them into a centroid to summarize them.
By clustering the set of vulnerable scopes into a smaller set of centroids, we reduce the dimensionality of the feature space and make it easier for the model to perform matching during inference. Additionally, the use of centroids ensures that similar vulnerable scopes are mapped to the same location in the feature space.
During training, we minimize the Wasserstein distance~\cite{feydy2019interpolating} between the set of vulnerable scopes and the vulnerability codebook, which allows us to effectively cluster vulnerable scopes and learn the representative centroids in the codebook.
During inference, our model matches the input program against all centroids in the learned vulnerability codebook.
By examining all the vulnerability patterns in the codebook, the matching process enables a thorough search for potential vulnerabilities. This explicit matching method supports the identification of specific vulnerability patterns and their associated statements, providing a comprehensive approach to identifying vulnerabilities.
We name this model \ourapp, a function and statement-level vulnerability identification approach via \underline{opti}mal transport quantization and vulnerability \underline{match}ing.}

In summary, our work presents several contributions: (i) an innovative vulnerability-matching DL framework utilizing optimal transport and vector quantization for function and statement-level vulnerability detection (VD); (ii) a novel statement embedding approach using recurrent neural networks (RNNs); and (iii) a thorough evaluation of our proposed method compared to other DL-based vulnerability prediction techniques on a large benchmark dataset of real-world vulnerabilities.

\vspace{-3mm}
\section{Related Work}
\vspace{-2mm}
Researchers have proposed various deep learning-based vulnerability detections (VDs) such as convolutional neural networks (CNNs)~\cite{russell2018automated}, recurrent neural networks (RNNs)~\cite{li2018vuldeepecker,vannguyen2019dan,van-nguyen-dual-dan-2020}, graph neural networks (GNNs)~\cite{zhou2019devign,chakraborty2021deep,li2021vulnerability,hin2022linevd,nguyen2022regvd,ding2022velvet}, and pre-trained transformers~\cite{feng2020codebert,guographcodebert,fu2022linevul,ding2022velvet}. RNN-based methods~\cite{russell2018automated,li2018vuldeepecker,li2021sysevr} have been shown more accurate than program analysis tools such as Checkmarx~\cite{checkmarx} and RATS~\cite{rats} to predict function-level vulnerabilities.
However, RNNs face difficulty in capturing long-term dependencies in long sequences as the model's sequential nature may result in the loss of earlier sequence information.
Furthermore, function-level predictions lack the required granularity to accurately identify the root causes of vulnerabilities.
Thus, researchers have proposed transformer-based methods that predict statement-level vulnerabilities and capture long-term dependencies~\cite{ding2022velvet,fu2022linevul} while ICVH~\cite{nguyen2021information} leverages bidirectional RNNs with information theory to detect statement-level vulnerabilities.
On the other hand, Zhou~\ea~\cite{zhou2019devign} embed the abstract syntax tree (AST), control flow graph (CFG), and data flow graph (DFG) for a code function and learn the graph representations for function-level predictions. Nguyen~\ea~\cite{nguyen2022regvd} proposed constructing a code graph as a flat sequence for function-level predictions. Hin~\ea~\cite{hin2022linevd} constructed program dependency graphs (PDGs) for functions and predicted statement-level vulnerabilities.

In contrast to the above methods, we propose a deep learning-based vulnerability matching method inspired by the principles of program analysis security tools. Specifically, we gather a group of vulnerability patterns from the training set and develop a vulnerability codebook using optimal transport~\cite{feydy2019interpolating} and vector quantization~\cite{van2017neural}. Our goal is to detect statements caused the vulnerabilities by matching functions with the representative patterns we have learned in our codebook.

\vspace{-3mm}
\section{Approach}
\vspace{-3mm}
Deep learning (DL) models have been proving their abilities in capturing vulnerabilities more accurately than program analysis tools using implicit vulnerability patterns learned from the training data set~\cite{croft2021empirical,fu2022linevul}.
However, in real-world source code data sets, common vulnerable scopes would be written in different styles (e.g., variable naming conventions) and appear at different spatial locations in different vulnerable sections (i.e., functions or programs)~\cite{nguyen2021information}.
Existing DL-based VD approaches often fail to consider the common vulnerable scopes (which could be clustered into patterns) that exist in vulnerable functions or programs during both training and inference, instead relying on implicit learning through supervised learning. To address this limitation, we propose a novel DL framework that integrates vulnerable scopes into centroids via a vulnerability codebook.
The example in Figure~\ref{fig:example_code} demonstrates that the two vulnerable functions have the same vulnerable scope, consisting of two vulnerable statements, but are presented in different variable names and spatial locations. To overcome this issue, we group these vulnerable scopes with the same pattern and quantize them into a codebook containing representative vulnerability centroids that can represent a set of similar scopes. This codebook is then used to facilitate vulnerability matching during the inference phase, effectively addressing the lack of consideration for vulnerable scopes in existing DL-based VD approaches.

\begin{figure*}[h!]
\centering
\vspace{-3mm}
\includegraphics[width=0.98\textwidth]{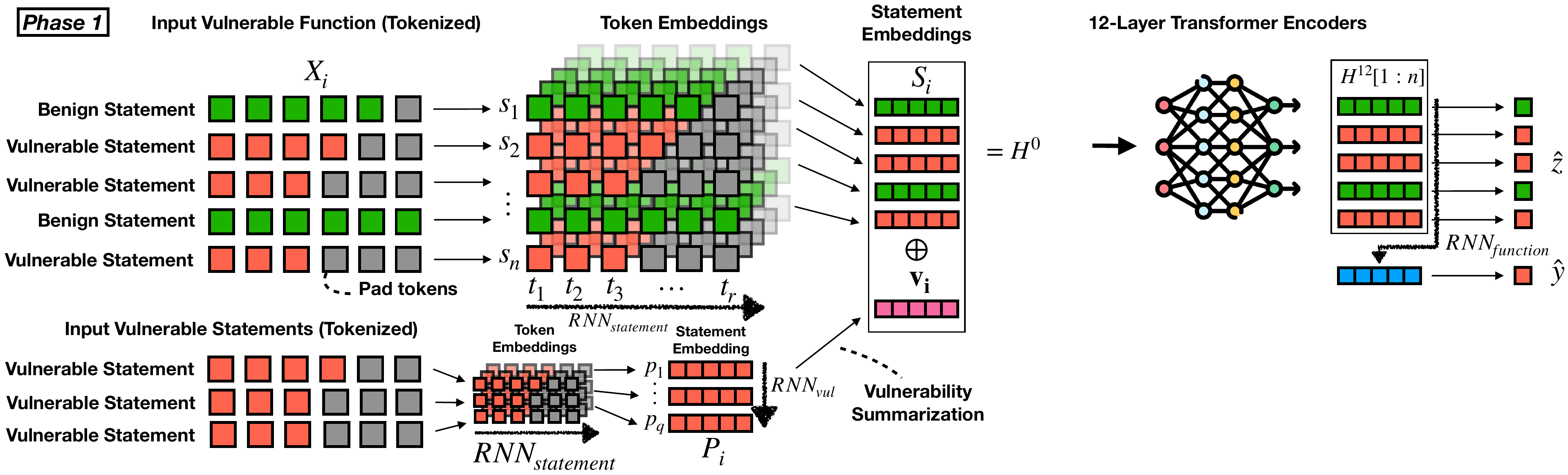}
\caption[Caption]{The overview of the warm-up phase in our approach. We tokenize each statement in a vulnerable function (i.e., $X_{i}$) followed by an embedding layer to map each token into a vector. We use $RNN_{statement}$ to summarize the token embeddings and get the statement embedding ($S_{i}$, $P_i$). For benign functions, $P_{i}$ is replaced by a special learnable embedding, $P_{benign}$. Additionally, we use $RNN_{vul}$ to summarize vulnerable statement embeddings $P_{i}$ to a vector $\mathbf{v_{i}}$ that represents the vulnerable scope. We concatenate $S_{i}$ and $\mathbf{v_{i}}$ as the input to transformer encoders to consider vulnerable scopes that arise in the function and align with our vulnerability matching process introduced in Section~\ref{sec:vul_matching}. We select the statement embeddings output from the last encoder, i.e., $H^{12}[1:n]$. Each statement embedding vector is mapped to a probability as statement-level predictions, the function-level prediction is obtained by summarising $H^{12}[1:n]$ to a vector using an $RNN_{function}$ and mapping it to a probability.}
\label{fig:overview_phase_1}
\vspace{-4mm}
\end{figure*}

In general, our approach consists of two phases.
The warm-up phase illustrated in Figure~\ref{fig:overview_phase_1} aims to gradually adjust the model parameters, with the goal of improving the representation of embeddings for input programs and vulnerable scopes.
The main training phase is illustrated in Figure~\ref{fig:overview_phase_2}.  
The yellow section on the left shows how we construct and learn our vulnerability codebook from vulnerable scopes in our training data using optimal transport. The grey section on the right shows how to utilize the codebook during training, which matches functions with the learned vulnerability centroids in the codebook, allowing us to identify and highlight the statements that caused the vulnerabilities.
Below, we first formulate our problem by defining common notations followed by how we map textual source code to vector space and warm up the embeddings.
We then introduce the motivation and method on why and how to learn a DL framework to achieve vulnerability matching.

\vspace{-2mm}
\subsection{Problem Statement}
\label{sec:problem_statement}
Let us consider a dataset of $N$ functions in the form of source code.
The data set includes both vulnerable and benign functions, where the function-level and statement-level ground truths have been labeled by security experts.
We denote a function as a set of code statements, $X_{i} = [\mathbf{x}_{1}, ..., \mathbf{x}_{n}]$, where $n$ is the max number of statements we consider in a function.
Let a sample of data be $\bigl\{(X_{i}, y_{i}, \mathbf{z}_{i}) : X_{i} \in \mathcal{X}, y_{i} \in \mathcal{Y}, \mathbf{z}_{i} \in \mathcal{Z}, i \in \{1, 2, ..., N\}\}$, where $\mathcal{X}$ denotes a set of code functions, $\mathcal{Y} = \{0, 1\}$ with 1 represents vulnerable function and 0 otherwise, and $\mathcal{Z}=\{0,1\}^{n}$ denotes a set of binary vectors with 1 represents vulnerable code statement and 0 otherwise.
Our objective is to identify the vulnerability on both \textit{function and statement levels}.
We formulate the identification of vulnerable functions as a binary classification problem and the identification of vulnerable statements as a multi-label classification problem.
Given a function $X_{i}$, we first input to a statement embedding layer ($SEMB$) to obtain statement embeddings, namely $S_{i}$ and $P_{i}$, as specified in Equation~\ref{eq:statement_emb} (refer to Section~\ref{statement_embeddings} for the embedding details).
$S_i \in \mathbb{R}^{n \times d}$ is the d-dimensional statement embedding vectors for $X_{i}$.
Prior studies have found that in a vulnerable function, there are code statements associated with the vulnerabilities (i.e., vulnerable scopes)~\cite{nguyen2021information}.
\revise{Let us denote $X_i^{vul}$ as a set of all vulnerable statements in a vulnerable function.
To explicitly capture vulnerable scopes, we extract $X_i^{vul}$ from the vulnerable function and encode those statements using d-dimensional statement embeddings as $P_i \in \mathbb{R}^{q \times d}$.} $q$ is the maximum number of vulnerable statements we consider in a vulnerable function and we set $q=12$ by applying truncation and padding because 95\% of vulnerable functions in our data have less than 12 vulnerable statements. Note that for each benign function without any vulnerable statements, we leverage a special learnable embedding denoted as $P_{benign} \in \mathbb{R}^{q \times d}$ to represent $P_i$.
In addition, we apply an RNN layer ($RNN_{vul}$) to summarize $P_i$ into a flat vector denoted as $\mathbf{v_i} \in \mathbb{R}^{d}$, which can facilitate the learning of our vulnerability codebook introduced in Section~\ref{sec:optimal_transport}.
Let us denote a stack of transformer encoders as $\mathcal{F}$, we concatenate $S_{i}$, $\mathbf{v_i}$, and feed them into transformer encoders as $\mathcal{F} (S_i, \mathbf{v_i})$.
We then make function and statement-level predictions based on the output of $\mathcal{F}$.
The mapping from $X_{i}$ to $y_{i}$ and $z_{i}$ is learned by minimizing the cross-entropy loss function, denoted by $\mathcal{L}(\cdot)$, as follows:
\begin{equation}
\label{eq:loss_function}
    min \frac{1}{N} \sum_{i=1}^{N} \Bigl[ \mathcal{L}_{function}\Bigl( \mathcal{F} (S_i, \mathbf{v_i}), y_{i} | X_{i} \Bigl) + \mathcal{L}_{statement}\Bigl( \mathcal{F} (S_i, \mathbf{v_i}), \mathbf{z}_{i} | X_{i} \Bigl) \Bigl]
\end{equation}

\subsection{Statement Embedding Using RNN}
\label{statement_embeddings}
Figure~\ref{fig:overview_phase_1} depicts the forward passes involved in our warm-up step to adjust the embeddings for statements and vulnerable scopes. 
We now introduce our motivations and method to embed statements and vulnerable scopes.
Large language models (LLMs) pre-trained for source code have been shown effective to predict vulnerabilities~\cite{feng2020codebert,guographcodebert,hin2022linevd,fu2022linevul}.
However, those LLMs leverage token embeddings that only preserve 512 tokens (tokenized by the byte pair encoding (BPE) algorithm~\cite{sennrich2016neural}) for each input function while extra tokens need to be truncated.
This could lead to information loss for long functions with more than 512 tokens.
To address this limitation, we propose the statement embedding layer $SEMB$ to encode a function (e.g., $X_i$) as a set of statement embeddings:
\begin{equation}
\label{eq:statement_emb}
    S_i = SEMB(X_i),\hspace{1mm} P_i = SEMB({X_i^{vul}}),\hspace{1mm}\textrm{where}\hspace{1mm} X_i, X_i^{vul}\in \mathcal{X}
\end{equation}
Given $X_i = [\mathbf{x}_{1}, ..., \mathbf{x}_{n}]$, we use BPE to tokenize $\mathbf{x}_{j}$ to a list of tokens, $[t_{1}, ..., t_{r}]$, where $r$ is the number of tokens we consider in a code statement.
We then obtain a token embedding for each $t_{j}$ using an embedding layer $E \in \mathbb{R}^{v \times d}$ where $v$ is the vocab size of the tokenizer.
This results in a token embedding matrix $\bar{S_i} \in \mathbb{R}^{n \times r \times d}$ for all statements in $X_i$.
Similarly, we obtain token embeddings of vulnerable statements $X_i^{vul}$ as $\bar{P_i} \in \mathbb{R}^{q \times r \times d}$ to represent a vulnerable scope. We apply truncation and padding to make $q$ a constant for each vulnerable function.

With $n=155$ and $r=20$ (see Section~\ref{sec:param_setting}), we can process 3,100 tokens per function, which is six times more than the 512 tokens. Our statement embedding method provides a more complete representation of code functions compared to the token embedding method. Specifically, our method can fully represent 99\% of the functions in our dataset that have less than 2,700 tokens, while the token embedding method can only fully represent around 85\% of the functions that have less than 500 tokens.
Table~\ref{tab:main_results} shows that our statement embedding method results in a 33\% and 32\% enhancement in the performance of CodeBERT and CodeGPT models for statement-level predictions.

Previous studies such as Sentence-BERT~\cite{reimers2019sentence} leverage max or mean pooling to aggregate token embeddings. 
The max pooling would lead to information loss since it considers the maximum token embedding for each statement, discarding all other token embeddings in the sequence. While the mean pooling considers all token embeddings, it treats all the token embeddings equally regardless of their importance or relevance to the statement they belong where the prominent token features could be disregarded.
In contrast, we propose to learn an RNN~\cite{cho2014properties} with $r$ (max number of tokens in each statement) time steps to aggregate the token embeddings and obtain statement embeddings as below:
\begin{equation}
\label{eq:gru_aggregation_1}
    S_i[j] = RNN_{statement}(\bar{S}[j, :, :]), \forall_{j} \in \{1, \ldots, n\}
\end{equation}
\begin{equation}
\label{eq:gru_aggregation_2}
    P_i[j] = RNN_{statement}(\bar{P}[j, :, :]), \forall_{j} \in \{1, \ldots, q\}
\end{equation}
\revise{To acquire the $j^{th}$ statement embedding for $S_i$ and $P_i$, we summarize the token embeddings of length $r$ using $RNN_{statement}$. Following the convention of Python lists, we represent the $j^{th}$ statement embeddings as $S_i[j]$.}
While mean or max pooling operations are not learnable, the $RNN_{statement}$ layer allows us to learn to pool token embeddings in each statement into a statement embedding vector while preserving prominent token features and mitigating the potential information loss.
Finally, we use $RNN_{vul}$ to summarize our vulnerable scope $P_i$ into a flat vector $\mathbf{v_i}$ (see Section~\ref{sec:scope_collection} for more details).
\vspace{-2mm}

\subsection{Training of Warm-Up Phase}
\label{sec:warm_up_training}
To consider the statement embeddings and the vulnerable scope of $X_{i}$, we concatenate $S_i$ and $\mathbf{v_i}$ to obtain the input to transformer encoders as $H^{0} = S_i \oplus \mathbf{v_i}$.
We select the statement embeddings output from the trail encoder, i.e., $H^{12}[1:n]$ where the $\mathbf{v_i}$ embedding is omitted.
We provide details of the transformer self-attention operation in Appendix~\ref{app:transformer_encoders}.
We use $RNN_{function}$ with $n$ time steps to summarize statement embeddings into a vector and map it to the function-level prediction $\hat{y_{i}} \in [0,1]$ as follows:
\begin{equation}
\label{eq:function_prediction}
    \hat{y_{i}} = \sigma\Bigl(drop\bigl(tanh(drop(RNN(H_{1:n}^{12}))W^{G})\bigl)W^{U}\Bigl)
\end{equation}
where $W^{G}\in\mathbb{R}^{d \times d}$ and $W^{U}\in\mathbb{R}^{d \times 1}$ are model parameters, $drop$ is a dropout layer, and $\sigma$ is a sigmoid function.
We map statement embeddings to a statement-level prediction $\hat{z_{i}} = [\hat{z_{i}}^{1}, \ldots, \hat{{z_i}}^{n}] \in [0,1]^n$ via:
\begin{equation}
\label{eq:statement_prediction}
    \hat{z_{i}} = \sigma\Bigl(drop\bigl(tanh(drop(H_{1:n}^{12})W^{I})\bigl)W^{J}\Bigl)
\end{equation}
where $W^{I} \in \mathbb{R}^{d \times d}$ and $W^{J} \in \mathbb{R}^{d \times 1}$ are model parameters, and $\sigma$ is a sigmoid function.

\begin{figure*}[h!]
\centering
\includegraphics[width=0.98\textwidth]{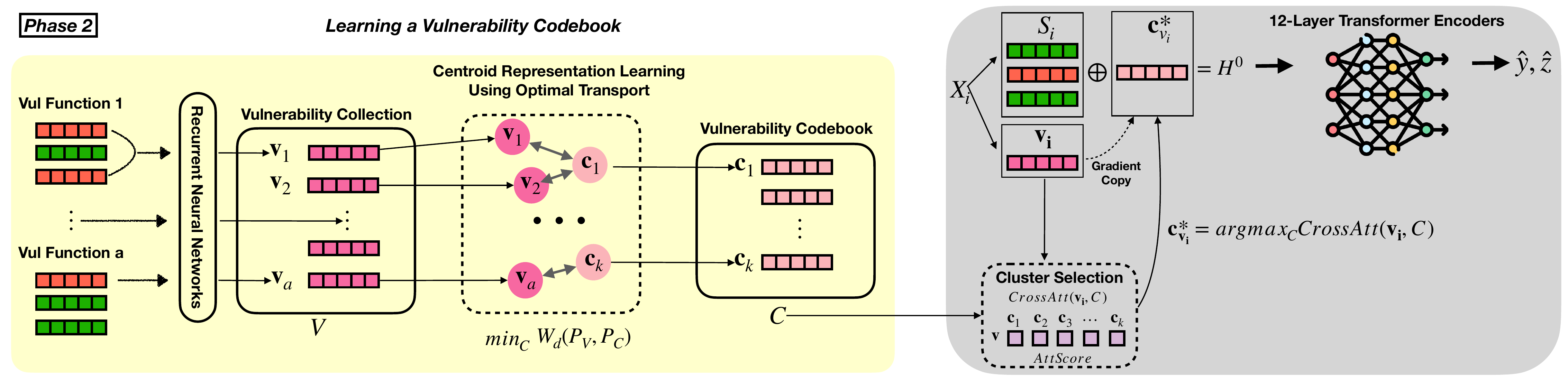}
\caption[Caption]{The overview of the main training phase in our approach. We introduce how to learn our vulnerability codebook on the left. We first collect a set of vulnerable statement embeddings from our training data. We then use $RNN_{vul}$ to pool a set of statement embeddings from each vulnerable function, forming a vulnerable scope represented by a vector $\mathbf{v_i}$. The set of these scopes forms our vulnerability collection $V=\{\mathbf{v}_{1}, \ldots, \mathbf{v}_{a}\}$. Next, we learn vulnerability centroids $\mathbf{c}_{j}$ using the Wasserstein distance metric to create a more compact vulnerability codebook $C=\{\mathbf{c}_{1}, \ldots, \mathbf{c}_{k}\}$, where each centroid represents a group of vulnerable scopes. During training, we minimize the Wasserstein distance between each $\mathbf{v_i}$ and its corresponding vulnerability centroid $\mathbf{c}_{\mathbf{v_i}}^{*}$. We illustrate this main training phase on the right side which is the same as our warm-up phase except that we concatenate $S_i$ and $\mathbf{c}_{\mathbf{v_i}}^{*}$ to obtain $H^{0}$ as detailed in Section~\ref{sec:main_training_phase}. To overcome the non-differentiability of the $argmax$ operation in the networks, we copy the gradients from $\mathbf{v}$ to $\mathbf{c}_{\mathbf{v_i}}^{*}$ to learn the statement embedding and pattern summarization RNNs for vulnerability patterns.}
\label{fig:overview_phase_2}
\vspace{-4mm}
\end{figure*}
\subsection{Vulnerability Codebook and Subsequent Main Training Phase}
Our model parameters are now warmed up to embed statements and vulnerable scopes.
Our objective is to achieve vulnerability matching using trainable vulnerability centroids.
In the following, we outline our motivations and approach for creating, training, and employing our vulnerability codebook during the primary training phase.

\subsubsection{Collect vulnerable scopes from Vulnerable Functions}
\label{sec:scope_collection}
To exploit and capture common vulnerable scopes in source code, we aim to learn a \textit{vulnerability codebook} containing representative centroids that group vulnerable scopes sharing the same pattern.
Unlike those patterns in program analysis tools, our vulnerability centroids are represented in vectors to conform with DL models, whose representation is adjustable during training, enabling the model to recognize typical vulnerability patterns that may occur at various spatial locations within a vulnerable function.

\revise{Given training data consisting of $a$ vulnerability functions, we extract vulnerable statements to form vulnerable scopes for each function as presented in the very first left part of Figure~\ref{fig:overview_phase_2}.
To simplify the process of building our vulnerability codebook introduced in Section~\ref{sec:optimal_transport}, we take two steps. First, we use $RNN_{vul}$ to summarize our vulnerable scopes into flat vectors. Then, we reduce the dimensionality of these vectors. This enables us to easily group them into vulnerability centroids and construct our vulnerability codebook.
We have denoted our vulnerable scope as $\mathbf{v_i}$ in Section~\ref{sec:problem_statement}.
$\mathbf{v_i}$ is obtained by applying $RNN_{statement}$ and $RNN_{vul}$ to get the vulnerable statement embeddings and condense them into a flat vector.  
To reduce the dimensionality of $\mathbf{\mathbf{v_i}}$, we linearly project the d-dimensional vector to the h-dimensional and normalize it as $\mathbf{v_i} = LN(\mathbf{v_i} \cdot W^{F})$ where $W^{F} \in \mathbb{R}^{d \times h}$ is model parameters and $LN$ is layer normalization.
We then accumulate each $\mathbf{v_i}$ extracted from vulnerable functions to form a \textit{vulnerability collection} denoted as $V \in \mathbb{R}^{a \times h}$ where $a$ is the total number of vulnerable functions in our training data.}

\subsubsection{Learn to Transport Vulnerable Scopes to Vulnerability Centroids in Codebook}
\label{sec:optimal_transport}
However, $V$ may consist of repeated or similar vulnerable scopes.
Additionally, the huge collection size of $V$ will also require many computing resources during inference since we need to match each function with $a$ number of scopes (in our training data, $a=6,361$).
To address such issues, we propose to learn a vulnerability codebook denoted as $C = [\mathbf{c}_{1}, ..., \mathbf{c}_{k}]$ where $\mathbf{c}_{i} \in \mathbb{R}^{h}$ is a vulnerability centroid.
Intuitively, this codebook integrates similar vulnerable scopes and forms common vulnerability patterns.
In particular, we reduce the 6,361 number of $\mathbf{v}$ vectors in our vulnerability collection to 150 vulnerability centroids in our codebook.

To ensure that vulnerability centroids can represent a group of similar vulnerable scopes, we leverage the optimal transport theory to transfer vulnerability patterns to their corresponding vulnerability centroid.
We minimize the Wasserstein distance~\cite{Villani2008OptimalTO} using the Sinkhorn approximation~\cite{sinkhorn} between our vulnerability collection and codebook.
Consequently, the vulnerable scopes and their respective vulnerability centroids will converge towards each other.
\revise{Ultimately, our codebook will comprise vulnerability centroids acting as representative patterns that symbolize different sets of vulnerability scopes.}
This allows us to aggregate similar vulnerability patterns based on Euclidean distance.
We summarize the process as follows:
\begin{equation}
\label{eq:wd_1}
    min_{C}\hspace{1mm} W_{d}(P_{V}, P_{C}), \hspace{2mm} \textrm{where} \hspace{2mm} P_{V} = \frac{1}{a}\sum_{i=1}^{a} \delta_{\mathbf{v}_{i}} \hspace{2mm}\textrm{and}\hspace{2mm} P_{C} = \frac{1}{a}\sum_{j=1}^{a} \delta_{\mathbf{c}_{j}}
\end{equation}
where $W_{d}$ is the Wasserstein distance \cite{Villani2008OptimalTO} and $\delta$ represents the Dirac delta distribution. According to the clustering view of optimal transport \cite{tidot, ho2017multilevel}, when minimizing $min_{C}\hspace{1mm} W_{d}(P_{V}, P_{C})$, the set of codebooks $C$ will become the centroids of the clusters formed by $V$. This clustering approach ensures that similar vulnerable scopes potentially sharing the same vulnerability pattern are grouped together, leading to a quantized vulnerability codebook that is more concise and effective.
We randomly initialize the embedding space of our vulnerability codebook as $C = [\mathbf{c}_{1}, ..., \mathbf{c}_{k}]$ with $k$ number of clusters.


\subsubsection{Main Training Phase}
\label{sec:main_training_phase}
The right part of Figure~\ref{fig:overview_phase_2} highlighted in grey summarizes our main training phase.
We load the model parameters warmed up in our previous phase. 
By employing the same statement embedding methodology introduced in Section~\ref{statement_embeddings}, we obtain the statement embeddings $S_i$ and a summarized vulnerable scope vector $\mathbf{v_i}$ for the input function $X_i$.

\revise{Instead of concatenating $S_i$ with $\mathbf{v_i}$, we employ a cluster selection process to map the vulnerable scope $\mathbf{v_i}$ to its most similar vulnerability centroid (denoted as $\mathbf{c}_{\mathbf{v_i}}^{*} \in \mathbb{R}^{1 \times h}$) selected from our codebook.
By doing so, the model inherently develops an understanding of the vulnerability centroids stored in our vulnerability codebook, which are closely linked to vulnerable functions.
We utilize the cross-attention (see Appendix~\ref{app:cross_attention}) between the vulnerable scope and the codebook and determine the vulnerability centroid for $\mathbf{v_i}$ as $\mathbf{c}_{\mathbf{v_i}}^{*} = argmax_{C} CrossAtt(\mathbf{v_i}, C)$.
The $argmax$ function selects the index of the vulnerability centroid with the highest attention score, which corresponds to the closest vector to $\mathbf{v_i}$ in terms of similarity.
We linearly project $\mathbf{c}_{\mathbf{v_i}}^{*}$ from the factorized h-dimension to the d-dimension to align with the dimension of our statement embeddings. 
Different from our warm-up phase where we concatenate $S_i$ with $\mathbf{v_i}$, we now concatenate $S_i$ with $\mathbf{c}_{\mathbf{v_i}}^{*}$ (the most similar centroid to the vulnerable scope $\mathbf{v_i}$).
Thus, the input to the encoders becomes $H^{0} = S_i \oplus \mathbf{c}_{\mathbf{v_i}}^{*}$.
The subsequent forward passes are the same as our warm-up phase described in Section~\ref{sec:warm_up_training}.}

Note that no real gradient is defined for $\mathbf{v_i}$ once we map it to a $\mathbf{c}_{\mathbf{v_i}}^{*}$ via an $argmax$ operation that causes the networks non-continuous and non-differentiable.
To let the networks which embed and summarize vulnerable statements be trainable via backpropagation, we follow the idea in VQ-VAE~\cite{van2017neural} which was shown effective for vector quantization.
We approximate the gradient similar to the straight-through estimator~\cite{bengio2013estimating} and copy gradients from summarized vulnerable scope $\mathbf{v_i}$ to selected vulnerability centroid $\mathbf{c}_{\mathbf{v_i}}^{*}$.
Below, we introduce how to leverage our learned codebook for vulnerability matching during inference.

\subsection{Vulnerability Identification Through Explicit Vulnerability Patterns Matching}
\label{sec:vul_matching}
\revise{Our approach utilizes vulnerable patterns that are often ignored by existing methods. By matching vulnerability centroids during inference, our approach enables us to fully harness the capabilities of DL models for vulnerability identification.
We first obtain d-dimensional statement embeddings $S_i$ from an input function $X_i$ as described in Section~\ref{statement_embeddings}.
For each vulnerability centroid $\mathbf{c_{j}}$ in our codebook, we linearly project $\mathbf{c_{j}}$ from h-dimensional to d-dimensional space and concatenate it with $S_{i}$ as $H_{j}^{0} = S_i \oplus \mathbf{c_{j}}$.
We then pass $H_{j}^{0}$ through transformer encoders ($\mathcal{F}$) to obtain function-level and statement-level vulnerability predictions, which is summarized as $P_i^{func}, P_i^{stmt} = \mathcal{F}(S_i, \mathbf{c_{j}}) \hspace{2mm} \forall_{j} \in \{1, \ldots, k\}$ where $P_{ij}^{func}\in [0,1]$ and $P_{ij}^{stmt}\in [0,1]^{n}$. Thus, we get $k$ (number of centroids in our codebook) function and statement-level predictions.
We use max pooling to pick the most prominent vulnerability-matching results as $\bar{P_{i}}^{func} = max_{k}P_i^{func}$ and predict if $X$ is a vulnerable function using a probability threshold of 0.5.
If $X$ is predicted as a benign function, we directly output a zero vector as the statement-level prediction.
Otherwise, we employ mean pooling to consider the prediction from each vulnerability centroid in our codebook as $\bar{P_{i}}^{stmt} = \frac{1}{k}\sum_{j=1}^{k}P_{ij}^{stmt}$ and predict if each statement is vulnerable using a probability threshold of 0.5.}
\vspace{-4mm}

\section{Experiments}
\vspace{-2mm}
\subsection{Experimental Dataset and Baseline Methods}
\vspace{-1mm}
\label{sec:dataset}
To identify vulnerabilities on function and statement levels, we select the Big-Vul data set created by Fan~\ea~\cite{fan2020ac} as it is one of the largest vulnerability data sets with statement-level vulnerability labels and has been used to assess statement-level vulnerability detection methods~\cite{hin2022linevd,fu2022linevul}. The data set was collected from 348 Github projects and consists of 188k C/C++ functions with 3,754 code vulnerabilities spanning 91 vulnerability types. The data distribution in our experiments resembles real-world scenarios, where the proportion of vulnerable to benign functions is 1:20. Our training data set comprises 6,361 vulnerability scopes before we group them into patterns in our codebook.

We compare our approach with (i) LLMs for code (i.e., CodeBERT~\cite{feng2020codebert} and GraphCodeBERT~\cite{guographcodebert}), (ii) Transformer-based VD (i.e., LineVul~\cite{fu2022linevul} and VELVET~\cite{ding2022velvet}), (iii) GNN-based VD (i.e., LineVD~\cite{hin2022linevd}, ReGVD~\cite{nguyen2022regvd}, and Devign~\cite{zhou2019devign}), (iv) RNN-based ICVH~\cite{nguyen2021information}, and (v) CNN-based TextCNN~\cite{chen2015convolutional}. More details of the baselines are provided in Appendix~\ref{sec:baselines}. 

\vspace{-1mm}
\subsection{Parameter Settings and Model Training}
\vspace{-1mm}
\label{sec:param_setting}
We split the data into 80\% for training, 10\% for validation, and 10\% for testing.
For both our approach and baselines, we consider $n=155$ statements in each function and $r=20$ tokens in each statement as the descriptive statistics of the whole data set suggest that 95\% of source code functions have less than 155 statements and 95\% of statements have less than 20 tokens. To initialize our transformer encoders, we make use of the pre-trained model provided by Wang~\ea~\cite{wang2021codet5}. This model has undergone pre-training through various denoising objectives associated with programming languages.
Details of the hyperparameter settings for our method in both phases are provided in Appendix~\ref{app:hyper_param}. In both training phases, we train our model through specific epochs and select the model that demonstrates the highest F1 score for statement-level prediction in the validation set.
The experiments were conducted on a Linux machine with an AMD Ryzen 9 5950X processor, 64 GB of RAM, and an NVIDIA RTX 3090 GPU.
\revise{The potential limitations imposed by our experimental setup are discussed in Appendix~\ref{app:discussion}.}

\vspace{-2mm}
\subsection{Main Results}
\vspace{-2mm}
We conduct our experiments several times and report the average numbers.
The experimental data set and baseline methods are detailed in Section~\ref{sec:dataset}.
We report accuracy (Acc), precision (Pre), recall (Re), and F1-score (F1) for function-level and statement-level vulnerability prediction tasks for a comprehensive evaluation of each approach.
This enables us to assess the models' performance on both positive and negative classes, regardless of the class imbalance between vulnerable and benign functions.
Note that the statement-level metrics are computed on the statement level instead of the function level to determine if each statement is correctly predicted.
The experimental results are shown in Table~\ref{tab:main_results}.
Our approach yields an improvement in function-level F1-score of 6\% to 65\% and an improvement in statement-level F1-score of 19\% to 71\%.These results highlight the effectiveness of our approach in accurately predicting vulnerabilities, both at the function and statement levels, outperforming all other state-of-the-art methods.
Furthermore, our RNN statement embedding method significantly enhances the performance of CodeBERT (30\% $\rightarrow$ 63\%) and CodeGPT (12\% $\rightarrow$ 44\%) in statement-level vulnerability prediction. This finding validates our intuition that the statement embeddings learned by our method can capture contextual information and locate statements associated with vulnerabilities more accurately than token embeddings.
\begin{table}[ht]
\centering
\caption{(Main Results) We compare our \ourapp~approach against other baseline methods and present results in percentage.}
\label{tab:main_results}

\resizebox{\textwidth}{!}{
\begin{tabular}{c|c|cccc|cccc}
\hline
\rowcolor[HTML]{DAE8FC} 
\cellcolor[HTML]{DAE8FC}                                  & \cellcolor[HTML]{DAE8FC}                                     & \multicolumn{4}{c|}{\cellcolor[HTML]{DAE8FC}\textbf{Function Level}} & \multicolumn{4}{c}{\cellcolor[HTML]{DAE8FC}\textbf{Statement Level}} \\ \cline{3-10} 
\rowcolor[HTML]{DAE8FC} 
\multirow{-2}{*}{\cellcolor[HTML]{DAE8FC}\textbf{Method}} & \multirow{-2}{*}{\cellcolor[HTML]{DAE8FC}\textbf{Embedding}} & \textbf{Acc}    & \textbf{Pre}   & \textbf{Re}     & \textbf{F1}     & \textbf{Acc}    & \textbf{Pre}    & \textbf{Re}     & \textbf{F1}    \\ \hline
\ourapp(ours)                                                      & Statement                                                    & \textbf{99.45}  & \textbf{97.66}  & \textbf{89.83}  & \textbf{93.58}  & \textbf{99.65}  & \textbf{86.8}  & \textbf{77.96}  & \textbf{82.14} \\ \hline
CodeBERT + our embedding                                                  & Statement                                                    & 98.91           & 92.15          & 82.89           & 87.28           & 99.19           & 59.39           & 67.84           & 63.33          \\
CodeBERT                                                  & Token                                                        & 98.75           & 93.9           & 77.27           & 84.78           & 96.89           & 19.29           & 63.54           & 29.6           \\ \hline
CodeGPT + our embedding                                                   & Statement                                                    & 98.95           & 91.25          & 84.81           & 87.91           & 98.23           & 32.54           & 67.34           & 43.88          \\
CodeGPT                                                   & Token                                                        & 95.69           & 56.18          & 19.02           & 28.42           & 98.48           & 14.4            & 9.7             & 11.6           \\ \hline
GraphCodeBERT                                             & Token                                                        & 95.51           & 50.11          & 27.03           & 35.12           & 96.94           & 10.56           & 26.34           & 15.08          \\ \hline
LineVul                                                   & Token                                                        & 98.61           & 89.25          & 78.47           & 83.51           & -               & -               & -               & -              \\
VELVET                                                    & Statement                                                    & 98.88           & 93.37          & 80.86           & 86.67           & 98.5            & 38.19           & 73.5            & 50.26          \\ \hline
LineVD                                                    & Statement                                                    & -               & -              & -               & -               & 95.19           & 27.1            & 53.3            & 36             \\
ReGVD                                                     & Token                                                        & 97.12           & 77.92          & 50.24           & 61.09           & -               & -               & -               & -              \\
Devign                                                    & Token                                                        & 96.9            & 72.29          & 50.24           & 59.28           & -               & -               & -               & -              \\ \hline
ICVH                                                      & Statement                                                    & 96.56           & 77.44          & 33.25           & 46.53           & 97.77           & 21.31           & 43.17           & 28.53          \\
TextCNN                                                   & Statement                                                    & 95.95           & 62.31          & 25.12           & 35.81           & 98.15           & 21.03           & 28.91           & 24.34          \\ \hline
\end{tabular}
}
\vspace{-5mm}
\end{table}

\vspace{-2mm}
\subsection{Ablation Study}
To assess the effectiveness of the proposed components in our \ourapp~approach, we conduct an ablation study. Specifically, we compare our RNN statement embedding method with mean or max pooling methods. Furthermore, we examine the impact of our vulnerability codebook and matching by comparing our approach with a variant that employs the same model architecture and pre-trained weights, but without using the vulnerability codebook and matching. Finally, we demonstrate the impact of the number of vulnerability centroids (i.e., $k$) on the performance of our approach.

The experimental results are shown in Table~\ref{tab:ablation_results}.
The utilization of mean or max pooling to summarize token embeddings into statement embeddings results in a slight decrease of 1.75\% and 0.45\% in function-level F1-score and 4.6\% and 4.12\% in statement-level F1-score, respectively, as compared to using an RNN. The results confirm the effectiveness of our RNN statement embedding method, indicating that it is more effective in summarizing token embeddings by retaining token features at each time step.
The performance significantly deteriorates by 33.58\% and 45.2\% for function and statement-level predictions when the vulnerability codebook and matching components are removed. This underscores the importance of these components in achieving high-performance levels. The results suggest that the vulnerability codebook plays a crucial role in our approach, which is responsible for retaining and leveraging the vulnerability patterns information present in vulnerable functions. This information is then utilized to identify vulnerable statements effectively during the vulnerability-matching inference.
The lower section of Table~\ref{tab:ablation_results} illustrates the impact of the number of vulnerability centroids on our approach. The results demonstrate that our approach attains favorable statement-level F1-scores for $k \in [100, 150, 200]$, and we set $k=150$ as it produces the optimal statement-level F1-score. Notably, $k$ represents a crucial factor, where a small value of $k$ (e.g., 50) may result in unsatisfactory performance due to the grouping of too many vulnerability patterns together, resulting in an inadequate representation of each pattern. Conversely, a large value of $k$ (e.g., 400) leads to a substantial embedding space of our codebook, making it challenging to update during the backward process. 
\begin{table}[ht]
\centering
\caption{(Ablation Results) We compare our proposed method to other variants to investigate the impact of the individual components. The metrics are reported as percentages.}
\label{tab:ablation_results}
\resizebox{\textwidth}{!}{
\begin{tabular}{c|cccc|cccc}
\hline
\rowcolor[HTML]{DAE8FC} 
\cellcolor[HTML]{DAE8FC}{\color[HTML]{000000} }                                          & \multicolumn{4}{c|}{\cellcolor[HTML]{DAE8FC}{\color[HTML]{000000} \textbf{Function Level}}}                                                                   & \multicolumn{4}{c}{\cellcolor[HTML]{DAE8FC}{\color[HTML]{000000} \textbf{Statement Level}}}                                                                  \\ \cline{2-9} 
\rowcolor[HTML]{DAE8FC} 
\multirow{-2}{*}{\cellcolor[HTML]{DAE8FC}{\color[HTML]{000000} \textbf{Method}}}         & {\color[HTML]{000000} \textbf{Acc}}   & {\color[HTML]{000000} \textbf{Pre}}   & {\color[HTML]{000000} \textbf{Re}}    & {\color[HTML]{000000} \textbf{F1}}    & {\color[HTML]{000000} \textbf{Acc}}   & {\color[HTML]{000000} \textbf{Pre}}  & {\color[HTML]{000000} \textbf{Re}}    & {\color[HTML]{000000} \textbf{F1}}    \\ \hline
{\color[HTML]{000000} \ourapp~(ours)}                                & {\color[HTML]{000000} \textbf{99.45}} & {\color[HTML]{000000} 97.66}          & {\color[HTML]{000000} 89.83}          & {\color[HTML]{000000} 93.58}          & {\color[HTML]{000000} \textbf{99.65}} & {\color[HTML]{000000} 86.8}          & {\color[HTML]{000000} 77.96}          & {\color[HTML]{000000} \textbf{82.14}} \\ \hline
{\color[HTML]{000000} w/o RNN embedding (mean pooling applied)}                          & {\color[HTML]{000000} 99.31}          & {\color[HTML]{000000} 98.49} & {\color[HTML]{000000} 86}             & {\color[HTML]{000000} 91.83}          & {\color[HTML]{000000} 99.59}          & {\color[HTML]{000000} \textbf{90.4}} & {\color[HTML]{000000} 67.89}          & {\color[HTML]{000000} 77.54}          \\
{\color[HTML]{000000} w/o RNN embedding (max pooling applied)}                           & {\color[HTML]{000000} 99.4}           & {\color[HTML]{000000} 96.53}          & {\color[HTML]{000000} 89.95}          & {\color[HTML]{000000} 93.13}          & {\color[HTML]{000000} 99.56}          & {\color[HTML]{000000} 79.7}          & {\color[HTML]{000000} 76.4}           & {\color[HTML]{000000} 78.02}          \\
{\color[HTML]{000000} w/o vulnerability codebook \& matching}                            & {\color[HTML]{000000} 94.81}          & {\color[HTML]{000000} 45.91}          & {\color[HTML]{000000} 86.6}           & {\color[HTML]{000000} 60}             & {\color[HTML]{000000} 98.19}          & {\color[HTML]{000000} 28.77}         & {\color[HTML]{000000} 51.57}          & {\color[HTML]{000000} 36.94}          \\ \hline
{\color[HTML]{000000} \ourapp~wt 50 vulnerability centroids}         & {\color[HTML]{000000} 85.9}           & {\color[HTML]{000000} 23.95}          & {\color[HTML]{000000} \textbf{98.21}} & {\color[HTML]{000000} 38.51}          & {\color[HTML]{000000} 95.5}           & {\color[HTML]{000000} 16.92}         & {\color[HTML]{000000} \textbf{86.13}} & {\color[HTML]{000000} 28.28}          \\
{\color[HTML]{000000} \ourapp~wt 100 vulnerability centroids}        & {\color[HTML]{000000} 99.38}          & {\color[HTML]{000000} 98.13}          & {\color[HTML]{000000} 87.92}          & {\color[HTML]{000000} 92.74}          & {\color[HTML]{000000} 99.64}          & {\color[HTML]{000000} 88.14}         & {\color[HTML]{000000} 74.98}          & {\color[HTML]{000000} 81.03}          \\
{\color[HTML]{000000} \ourapp~wt 150 vulnerability centroids (ours)} & {\color[HTML]{000000} \textbf{99.45}} & {\color[HTML]{000000} 97.66}          & {\color[HTML]{000000} 89.83}          & {\color[HTML]{000000} 93.58}          & {\color[HTML]{000000} \textbf{99.65}} & {\color[HTML]{000000} 86.8}          & {\color[HTML]{000000} 77.96}          & {\color[HTML]{000000} \textbf{82.14}} \\
{\color[HTML]{000000} \ourapp~wt 200 vulnerability centroids}        & {\color[HTML]{000000} \textbf{99.45}} & {\color[HTML]{000000} 96.69}          & {\color[HTML]{000000} 90.91}          & {\color[HTML]{000000} \textbf{93.71}} & {\color[HTML]{000000} 99.63}          & {\color[HTML]{000000} 83.44}         & {\color[HTML]{000000} 80.02}          & {\color[HTML]{000000} 81.69}          \\
{\color[HTML]{000000} \ourapp~wt 400 vulnerability centroids}        & {\color[HTML]{000000} 98.28}          & {\color[HTML]{000000} \textbf{99.05}}          & {\color[HTML]{000000} 62.32}          & {\color[HTML]{000000} 76.51}          & {\color[HTML]{000000} 99.54}          & {\color[HTML]{000000} 81.91}         & {\color[HTML]{000000} 70.47}          & {\color[HTML]{000000} 75.76}
\end{tabular}
}
\end{table}

\vspace{-6mm}
\section{Conclusion}
\vspace{-2mm}
This paper presents a novel vulnerability-matching method for function and statement-level vulnerability detection (VD). Our approach capitalizes on the vulnerability patterns present in vulnerable programs, which are typically overlooked in deep learning-based VD. To be specific, we collect vulnerability patterns from the training data and learn a more compact vulnerability codebook from the pattern collection using optimal transport (OT) and vector quantization. During inference, the codebook is utilized to match all learned patterns and detect potential vulnerabilities within a given program. Our comprehensive evaluation, conducted on over 188,000 real-world C/C++ functions, demonstrates that our method surpasses other competitive baseline techniques, while our ablation study confirms the soundness of our approach.

\bibliographystyle{plain}
\bibliography{reference}

\clearpage
\appendix
\section{Appendix}
\subsection{Self-Attention of Transformer Encoders}
\label{app:transformer_encoders}
Given the input $H^{0}$, we leverage 12 layers of transformer encoders to learn the representation of $x$ as follows:
\begin{equation}
\label{eq:xfmr_1}
    A^{t} = LN(MultiAttn(H^{t-1})) + H^{t-1}, t \in \{1, ..., 12\}
\end{equation}
\begin{equation}
\label{eq:xfmr_2}
    H^{t} = LN(FFN(A^{t}) + A^{t})
\end{equation}
where $A^{t}$ the self-attention output, $MultiAttn$ is a multi-head attention, $FFN$ is feed-forward neural networks and $LN$ is layer normalization.

\subsection{Cross Attention for Selecting Vulnerability Centroids}
\label{app:cross_attention}
\begin{equation}
\label{eq:cluster_assignment_1}
    Q = \mathbf{v_i} \cdot W^{Q}, K = C \cdot W^{K}, V = C \cdot W^{V} 
\end{equation}
\begin{equation}
\label{eq:cluster_assignment_2}
    AttScore = Drop(\psi(Q \cdot K^{T})), AttScore \in \mathbb{R}^{1 \times k}
\end{equation}
\begin{equation}
\label{eq:cluster_assignment_3}
    \mathbf{c}_{\mathbf{v_i}}^{*} = argmax(AttScore)
\end{equation}
where query states $Q$ is obtained by linearly projected $\mathbf{v_i}$ using model parameters $W^{Q} \in \mathbb{R}^{h \times h}$, key states $K$ and value states $V$ are obtained by linearly projected $C$ using model parameters $W^{K}, W^{V} \in \mathbb{R}^{h \times h}$. $\psi$ is a softmax function and $Drop$ is a dropout layer.
We use $argmax$ operation to obtain the codebook embedding index and map $\mathbf{v_i}$ to the corresponding $\mathbf{c}_{\mathbf{v_i}}^{*}$ having the maximum $AttScore$.

\subsection{Details of Baseline Methods}
\label{sec:baselines}
We compare our \ourapp~approach with LLMs pre-trained on source code data, state-of-the-art transformer-based, GNN-based, RNN-based, and CNN-based vulnerability detection (VD) approaches. We reproduce each baseline based on the code provided by the original authors.

\textbf{LLMs for code}:
We include CodeBERT~\cite{feng2020codebert}, CodeGPT~\cite{lu1codexglue}, and GraphCodeBERT~\cite{guographcodebert}. These models were pre-trained with token embeddings. We include an additional trial for CodeBERT and CodeGPT using our RNN statement embedding method. Note that the statement embedding is not compatible with GraphCodeBERT's data flow construction. 

\textbf{Transformer-based VD}:
LineVul~\cite{fu2022linevul} is designed to perform function-level prediction by leveraging a pre-trained transformer model. Although it can also provide statement-level predictions by interpreting and ranking the attention scores of the transformer, this approach is not suitable for the statement-level classification setting. To ensure a fair comparison, we only evaluate our approach against LineVul on the function level.
VELVET~\cite{ding2022velvet} is an ensemble method that leverages a vanilla transformer with GNNs.

\textbf{GNN-based VD}:
LineVD~\cite{hin2022linevd}, ReGVD~\cite{nguyen2022regvd}, and Devign~\cite{zhou2019devign} are GNN-based methods that learn the graph property of source code. Note that ReGVD and Devign only predict function-level vulnerabilities.

\textbf{RNN-based and CNN-based VD}:
ICVH~\cite{nguyen2021information} leverages Bi-RNN with information theory to detect statement-level vulnerabilities. ICVH was initially trained in the unsupervised setting for statement-level vulnerability prediction, but we found that it was not effective in our context. Therefore, we adopted the original ICVH architecture and added a cross-entropy loss to train ICVH in the supervised setting to achieve a fair comparison. On the other hand, TextCNN~\cite{chen2015convolutional} uses convolutional layers for sentence classification tasks.

\subsection{Hyper-Parameter Settings of Our \ourapp~Approach}
Table~\ref{tab:hyper_param} lists the hyper-parameter settings required to reproduce our approach. We have made our replication package available at \url{https://github.com/optimatch/optimatch}, which includes all the experimental scripts. We have included a comprehensive README file that contains all of the details needed to reproduce the experimental results demonstrated in this paper.

\label{app:hyper_param}
\begin{table}[ht]
\centering
\caption{The hyper-parameter settings of our \ourapp~approach.}
\label{tab:hyper_param}
\resizebox{\textwidth}{!}{
\begin{tabular}{c|ccccccccc}
\hline
\textbf{Phase} & \textbf{Optimizer} & \textbf{Scheduler}           & \textbf{LR} & \textbf{Grad Clip} & \textbf{Epochs} & \textbf{Stmt Len ($r$)} & \textbf{Max Num Stmt ($n$)} & \textbf{Num Centroids ($k$)} & \textbf{Batch} \\ \hline
Warm-up        & AdamW              & Linear (4,650 warm-up steps) & 1e-4        & 1.0                & 20              & 20 tokens                 & 155                            & -                         & 64             \\
Main           & AdamW              & Linear (4,650 warm-up steps) & 1e-4        & 1.0                & 20              & 20 tokens                 & 155                            & 150                       & 64            
\end{tabular}
}
\end{table}

\revise{\subsection{Discussion}
\label{app:discussion}
Inspired by program analysis tools for locating vulnerabilities based on predefined vulnerability patterns, we proposed our innovative vulnerability-matching deep-learning framework not only successfully utilizing optimal transport and vector quantization for function and statement-level vulnerability detection but also effectively leverage the information presented in vulnerable statements and patterns to enhance deep learning-based vulnerability detection. We found that the performance of our approach can be affected by the chosen number of vulnerability centroids ($k$) used in the codebook.  While our approach has shown promising results in detecting vulnerabilities in source code, the number of centroids is currently a hyperparameter that requires manual tuning. This could be a limitation of our approach when scaling up to larger datasets or more complex codebases, as it may not be feasible to manually optimize the number of centroids for each dataset. Thus, our future work should focus on developing automated methods for selecting the optimal number of centroids or incorporating more advanced techniques such as adaptive quantization to dynamically adjust the codebook dictionary during training. Nevertheless, we conduct an ablation study to reason the optimal solution in this paper.
We assessed the effectiveness of our method using the extensive Big-Vul dataset, which includes 188,000 C/C++ functions and 3,754 code vulnerabilities across 91 distinct CWE-IDs. This dataset is expected to be comprehensive and inclusive, containing a range of code patterns and vulnerabilities that are reflective of real-world scenarios. Nevertheless, the performance of our method and the comparison baselines may vary when tested on other datasets with different characteristics. We acknowledge this limitation and the potential bias in our findings.}


\end{document}